\documentclass[twoside,twocolumn,9pt]{article}
\usepackage{extsizes}
\usepackage[super,sort&compress,comma]{natbib}
\usepackage[version=3]{mhchem}
\usepackage[left=1.5cm, right=1.5cm, top=1.785cm, bottom=2.0cm]{geometry}
\usepackage{balance}
\usepackage{widetext}
\usepackage{times,mathptmx}
\usepackage{sectsty}
\usepackage{graphicx}
\usepackage{lastpage}
\usepackage[format=plain,justification=raggedright,singlelinecheck=false,font={stretch=1.125,small,sf},labelfont=bf,labelsep=space]{caption}
\usepackage{float}
\usepackage{fancyhdr}
\usepackage{fnpos}
\usepackage[english]{babel}
\usepackage{array}
\usepackage{droidsans}
\usepackage{charter}
\usepackage[T1]{fontenc}
\usepackage[usenames,dvipsnames]{xcolor}
\usepackage{setspace}
\usepackage[compact]{titlesec}
\usepackage{latexsym,amsmath}
\usepackage{pifont}
\usepackage{color}

\usepackage{epstopdf}

\newcommand{\cmark}{\ding{51}}%
\newcommand{\xmark}{\ding{55}}

\definecolor{cream}{RGB}{222,217,201}

\begin{document}

\pagestyle{fancy}
\thispagestyle{plain}
\fancypagestyle{plain}{

\fancyhead[C]{\includegraphics[width=18.5cm]{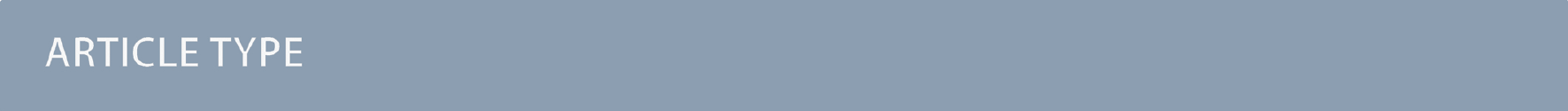}}
\fancyhead[L]{\hspace{0cm}\vspace{1.5cm}\includegraphics[height=30pt]{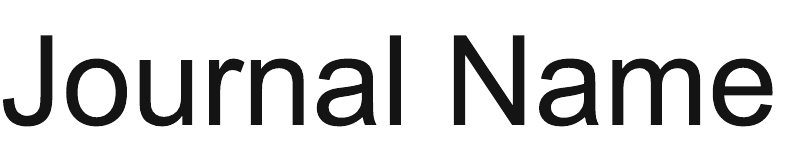}}
\fancyhead[R]{\hspace{0cm}\vspace{1.7cm}\includegraphics[height=55pt]{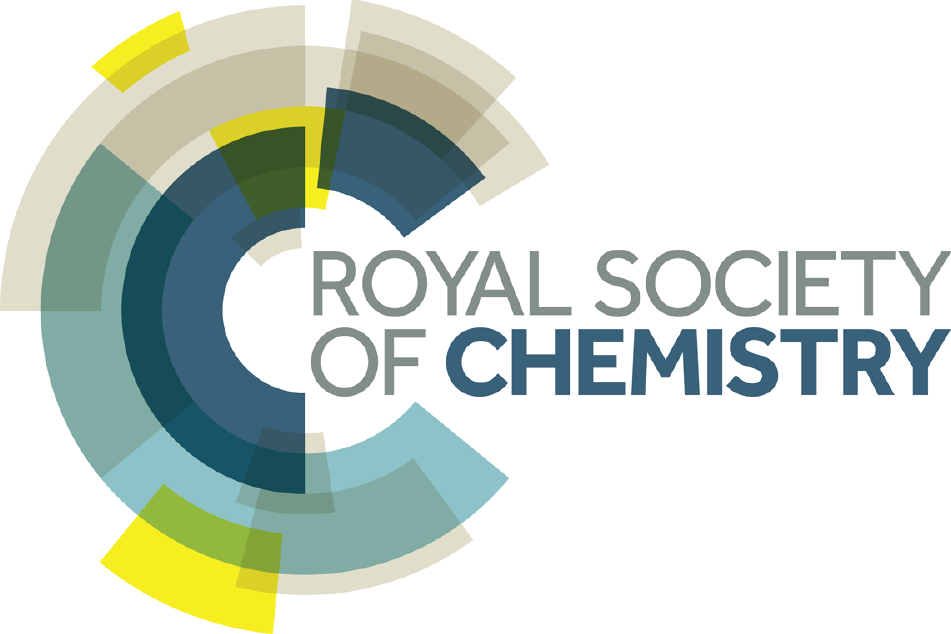}}
\renewcommand{\headrulewidth}{0pt}
}

\makeFNbottom
\makeatletter
\renewcommand\LARGE{\@setfontsize\LARGE{15pt}{17}}
\renewcommand\Large{\@setfontsize\Large{12pt}{14}}
\renewcommand\large{\@setfontsize\large{10pt}{12}}
\renewcommand\footnotesize{\@setfontsize\footnotesize{7pt}{10}}
\makeatother

\renewcommand{\thefootnote}{\fnsymbol{footnote}}
\renewcommand\footnoterule{\vspace*{1pt}%
\color{cream}\hrule width 3.5in height 0.4pt \color{black}\vspace*{5pt}}
\setcounter{secnumdepth}{5}

\makeatletter
\renewcommand\@biblabel[1]{#1}
\renewcommand\@makefntext[1]%
{\noindent\makebox[0pt][r]{\@thefnmark\,}#1}
\makeatother
\renewcommand{\figurename}{\small{Fig.}~}
\sectionfont{\sffamily\Large}
\subsectionfont{\normalsize}
\subsubsectionfont{\bf}
\setstretch{1.125} 
\setlength{\skip\footins}{0.8cm}
\setlength{\footnotesep}{0.25cm}
\setlength{\jot}{10pt}
\titlespacing*{\section}{0pt}{4pt}{4pt}
\titlespacing*{\subsection}{0pt}{15pt}{1pt}

\fancyfoot{}
\fancyfoot[LO,RE]{\vspace{-7.1pt}\includegraphics[height=9pt]{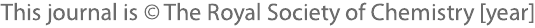}}
\fancyfoot[CO]{\vspace{-7.1pt}\hspace{13.2cm}\includegraphics{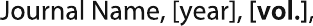}}
\fancyfoot[CE]{\vspace{-7.2pt}\hspace{-14.2cm}\includegraphics{RF}}
\fancyfoot[RO]{\footnotesize{\sffamily{1--\pageref{LastPage} ~\textbar  \hspace{2pt}\thepage}}}
\fancyfoot[LE]{\footnotesize{\sffamily{\thepage~\textbar\hspace{3.45cm} 1--\pageref{LastPage}}}}
\fancyhead{}
\renewcommand{\headrulewidth}{0pt}
\renewcommand{\footrulewidth}{0pt}
\setlength{\arrayrulewidth}{1pt}
\setlength{\columnsep}{6.5mm}
\setlength\bibsep{1pt}

\makeatletter
\newlength{\figrulesep}
\setlength{\figrulesep}{0.5\textfloatsep}

\newcommand{\topfigrule}{\vspace*{-1pt}%
\noindent{\color{cream}\rule[-\figrulesep]{\columnwidth}{1.5pt}} }

\newcommand{\botfigrule}{\vspace*{-2pt}%
\noindent{\color{cream}\rule[\figrulesep]{\columnwidth}{1.5pt}} }

\newcommand{\dblfigrule}{\vspace*{-1pt}%
\noindent{\color{cream}\rule[-\figrulesep]{\textwidth}{1.5pt}} }

\makeatother

\twocolumn[
  \begin{@twocolumnfalse}
\vspace{3cm}
\sffamily
\begin{tabular}{m{4.5cm} p{13.5cm} }

\includegraphics{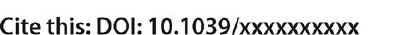} & \noindent\LARGE{\textbf{Density and structural anomalies in soft-repulsive dimeric fluids$^\dag$}} \\
\vspace{0.3cm} & \vspace{0.3cm} \\

& \noindent\large{Gianmarco Muna\'o$^{\ast}$\textit{$^{a}$} and Franz Saija\textit{$^{b}$}} \\

%

\includegraphics{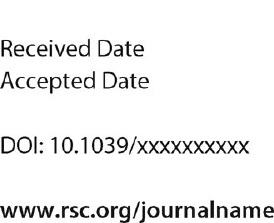} & \noindent\normalsize{
We report Monte Carlo results for the fluid structure of a system of dimeric
particles interacting via a core-softened potential. 
More specifically, dimers interact through a repulsive pair potential 
of inverse-power form, modified in such a way that the repulsion strength
is softened in a given range of distances.
The aim of such a study is to investigate how both the elongation of the dimers
and the softness of the potential affect some features of the model.
Our results show that the dimeric fluid exhibits both density and structural anomalies even
if the interaction is not characterized by two length scales.
Upon increasing the aspect ratio of the dimers, such anomalies are progressively 
hindered, with the structural anomaly surviving even after the disappearance of the density anomaly. 
These results shed light on the peculiar behaviour of molecular systems 
of non-spherical shape, showing how geometrical and interaction parameters 
play a fundamental role for the presence of anomalies.}

\end{tabular}

 \end{@twocolumnfalse} \vspace{0.6cm}

  ]

\renewcommand*\rmdefault{bch}\normalfont\upshape
\rmfamily
\section*{}
\vspace{-1cm}


\footnotetext{\textit{$^{a}$~Dipartimento di Scienza Matematiche e Informatiche, Scienze Fisiche e Scienze della Terra, Universit\'a degli Studi di Messina, Viale F. Stagno d'Alcontres 31, 98158 Messina, Italy. E-mail: gmunao@unime.it}}
\footnotetext{\textit{$^{b}$~CNR-IPCF, Viale F. Stagno d'Alcontres 37, 98158 Messina, Italy. Email: franz.saija@cnr.it}}


\section{Introduction}
The class of systems known as network-forming fluids, i.e. fluids characterized
by intermolecular bonds strictly dependent by the orientation, has always
captured great interest in the field of chemical physics. Some well-known
examples of such compounds, like 
water~\cite{Debenedetti:JPCM,Mishima:00,Soper:00,Poole:92}, 
carbon~\cite{Thiel:93}, 
phosphorous~\cite{Katayam:00} and silica~\cite{Lacks:00}, have
been deeply investigated because of their so-called anomalous behaviours,
including re-entrant melting, diffusion and density anomalies. 
Water, in particular, is still object of rather intense 
experimental~\cite{Dokter:05,Mallamace:12,Mallamace:13} and
theoretical~\cite{Liu:09,Poole:11,Franzese:12,Poole:13} 
investigations for its anomalous properties that deeply influence
its phase behaviour, even giving rise to a 
possible liquid-liquid critical 
point. In this context, the possibility to develop simple models able to 
reproduce such thermodynamics anomalies even via spherically symmetric 
potentials constitutes a fascinating challenge; in
the last years, a rich variety of models, based on some specific choices of the
interaction potential, have been proposed to this aim. One of the first 
studies is due to Jagla~\cite{Jagla:99}, who introduced a particular 
model for the intermolecular interaction 
by setting a short-range hard-core plus a linear repulsive shoulder at 
larger distances. The resulting phase diagram showed anomalous properties 
similar to those observed in water. The Jagla potential belongs to a more 
general class of intermolecular interactions called core-softened 
(CS) potentials:
firstly introduced by Hemmer and Stell in 1970~\cite{Stell:70} and later 
recovered by Debenedetti and coworkers~\cite{Borick:91}, these potentials
are characterized by the softening of the hard-core plus an attractive tail.
It has been observed~\cite{Stell:70} that such a softening may cause a 
second transition if a first already exists. After these preliminary works, a 
large variety of investigations has been carried out more recently to 
investigate the peculiar physical properties of CS potentials: 
more specifically, attention has been paid to thermodynamic anomalies of the
Hemmer-Stell potential~\cite{Scala:98,Scala:00} and of the Jagla 
potential~\cite{Wilding:02,Kumar:05,Gallo:12}, as well as to 
liquid-liquid phase 
transition~\cite{Malescio:01,Pelli:01,Pelli:04,Malescio:04,Wilding:06,franzese07,Hus:14} 
and waterlike anomalies in 
CS potentials~\cite{Scala:01,Barbosa:06,Malescio:08,Barbosa:EPJ,Barbosa:08,Barbosa:EPL,Buldyrev:09,Frenkel:09}. 
Also, CS potentials have 
been adopted to simulate, in a coarse-grain approach,
the phase behaviour of water~\cite{Hus-JCP} and 
alcohols~\cite{Urbic:14,Urbic:15}.

Such interactions are characterized by two competing, expanded and compact,
local arrangements of particles. This property, although
arising from simple isotropic interactions, effectively mimics the
behaviour of the much more complex network-forming fluids,
where loose and compact local structures arise from the continuous
formation and disruption of the dynamic network
originated by orientational bonds. Recently, the two-scale picture
as a requisite for anomalous behaviours has been challenged by some
studies showing that even a weak softening of the repulsive interparticle
interaction, though not able to yield two distinct length
scales, may nevertheless give origin to anomalous behaviours~\cite{Saija:09,Saija:10}.

In this study we investigate the fluid phase of a model
composed by dimeric particles interacting via a modified 
inverse-power potential (MIP)~\cite{Franz:JPCB} by performing extensive
Monte Carlo (MC) simulations in the canonical ensemble. In previous 
works~\cite{Franz:MolPhys,Franz:JPCB} it has been shown that a system of
spherical particles interacting through MIP can display typical water-like 
anomalies by varying carefully the softening parameter of the interparticle potential.
Such an approach allows one to follow the crossover from one-scale behaviour 
characterizing Lennard-Jones-like fluids to a two-scale behaviour typical of
CS potential systems. Here we
generalize this approach by considering symmetric dimers of variable aspect 
ratio, modeled as two partially fused spheres interacting with the spheres
of an other dimer via the
MIP potential. Starting with the monomeric case we progressively increase 
the aspect ratio with the aim to
investigate how some anomalous behaviours are
influenced by the red elongated shape of the molecules.
At the same time, the two different conditions corresponding to one and two
length scales of the potential are taken into account.
In a previous study by de Oliveira and coworkers~\cite{Barbosa:10} 
the authors studied a model consisting of dimeric
molecules which interact through a intermolecular continuous shoulder 
potential, but they limited their analysis to a single value 
of the aspect ratio.

The novelty of our study relies in the attempt 
to relate both the geometry 
of the system and the specific form of the interaction potential to the
onset of anomalies in molecular non-spherical fluids, with a possible 
application to a wide range of soft systems.

The paper is organized as follows: in the next section we provide
details of the model and the simulation approach. Results
are presented and discussed in the third section and conclusions follow in the
last section.

\begin{figure}
\begin{center}
\includegraphics[width=8.0cm,angle=0]{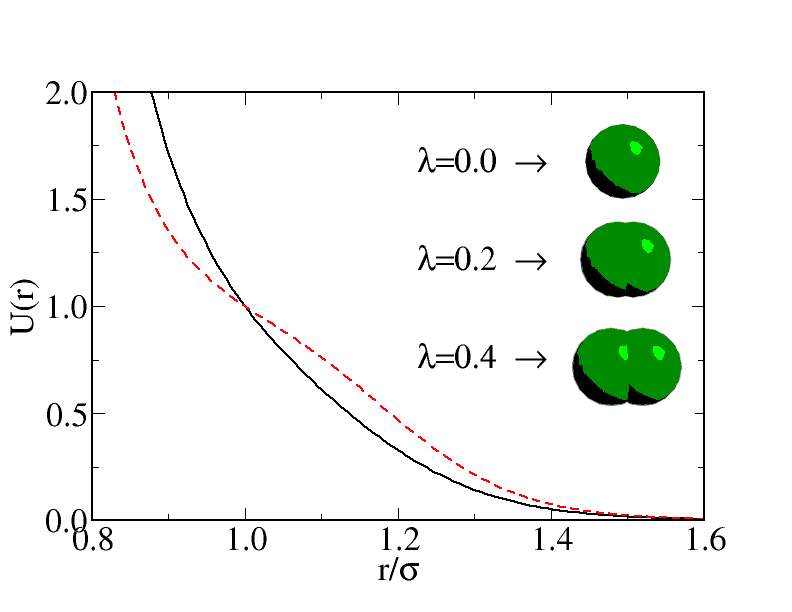} 
\caption{Site-site pair potential energy 
$U(r)$ at $\alpha=0.6$ (full line) and
$\alpha=0.8$ (dashed line). Cartoons of dimers upon increasing
$\lambda$ are also shown.}\label{fig:poten}
\end{center}
\end{figure}

\section{Model and simulations}
The sequence of models investigated in this work
is schematically depicted
in Fig.~\ref{fig:poten}: starting from two spheres totally overlapped 
we progressively increase the distance from their centres, 
decreasing the overlapping volume. 
In this way we obtain dimers comprised by two spherical 
particles rigidly bonded together.
If we indicate 
the aspect ratio $\lambda$ as the distance
between the two centers of the spheres, we move from $\lambda=0$ (total
overlap) to $\lambda=0.40$. 
In this range, each value attained by $\lambda$ is smaller than the radius 
$\sigma/2$ of the spheres constituting the dimers ($\sigma$ being the 
diameter). 
The interaction site-site potential
is set as:
\begin{equation}\label{eq:pot}
U(r)=\epsilon(\sigma/r)^{n(r)}
\end{equation}
where $\epsilon$ and $\sigma$ are the units of energy and length, respectively,
and
\begin{equation}\label{eq:nr}
n(r)=n_0 \{1-\alpha\exp[-b(1-r/\sigma)^2] \}
\end{equation}
In this equation, $\alpha$ is a real number between 0 and 1 and $b$ and $n_0$ 
are positive integer numbers. 
Also, we define reduced temperature, density
and pressure as, respectively, $T^*\equiv k_BT/\epsilon$, 
$\rho^*\equiv \rho\sigma^3$ and $P^*\equiv P\sigma^3/\epsilon$,
where $k_B$ is the Boltzmann constant.
\begin{figure}
\begin{center}
\begin{tabular}{c}
\includegraphics[width=8.0cm,angle=0]{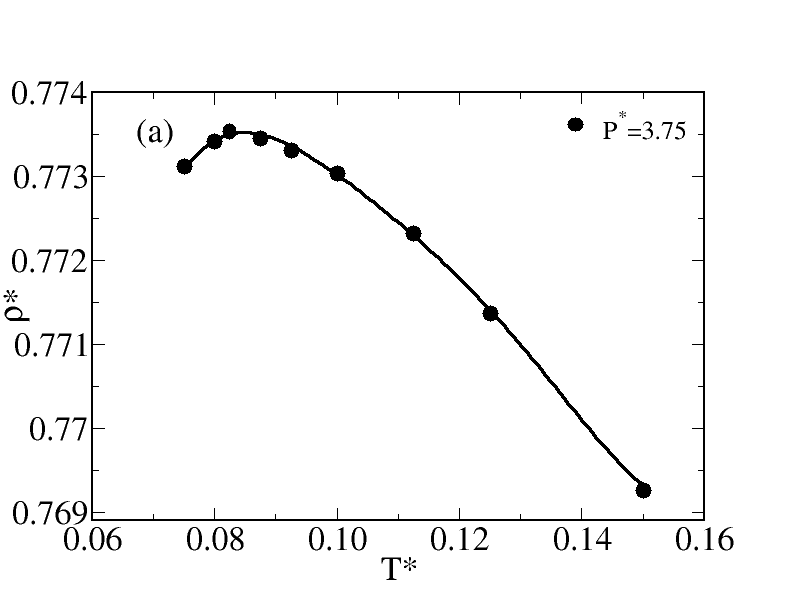} \\
\includegraphics[width=8.0cm,angle=0]{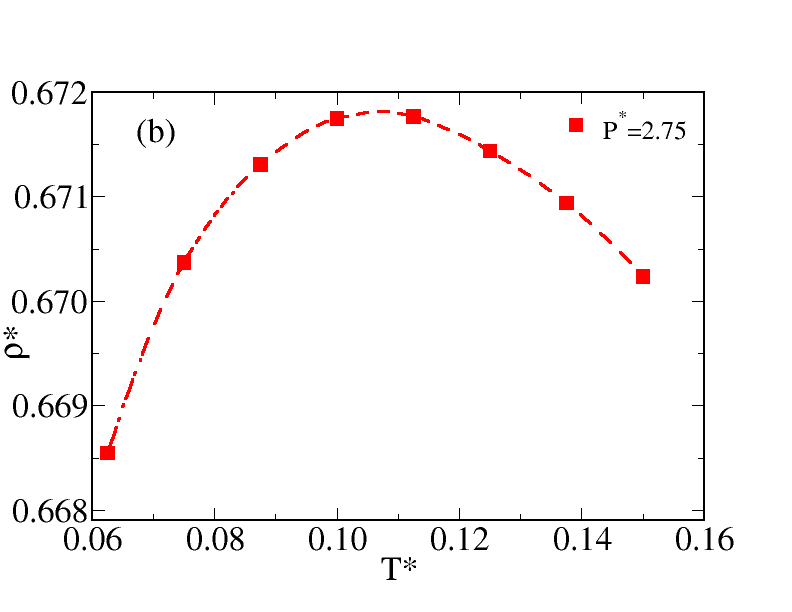} 
\end{tabular}
\caption{Density versus temperature at constant pressure for $\alpha=0.6$
and $\lambda=0.05$ (a) and for $\alpha=0.8$ and $\lambda=0.20$ (b).}
\label{fig:TMD}
\end{center}
\end{figure}

In Eqs.~\ref{eq:pot}-\ref{eq:nr}, introduced and investigated
in Refs.~\cite{Franz:JPCB,Franz:MolPhys}, $\alpha$ takes into account
the repulsion softening, while $b$ controls the width of the interval where
$n(r)$ is smaller than $n_0$. 
In the following, we keep fixed $b=5$ and
$n_0=12$ and investigate the two different cases corresponding to $\alpha=0.6$
and $\alpha=0.8$. The behaviour of $U(r)$ 
for these two cases is also reported in 
Fig.~\ref{fig:poten}: one can notice that for $\alpha=0.6$, the potential 
essentially follows an inverse-power law, whereas for $\alpha=0.8$ there is 
an inflection point with a change of the concavity. As 
demonstrated in a previous work on monomeric particles interacting via
the same potential of the model at issue~\cite{Franz:JPCB}, for $\alpha=0.6$
the potential exhibits one length scale, whereas for $\alpha=0.8$
two distinct, repulsive length scales emerge, 
giving rise to a competition between them.    
The condition $\lambda$ < $\sigma /2$ ensures that the 
effective global dimer-dimer interaction
preserves the same features of the site-site potential in terms of one or 
two length scales.
\begin{figure}
\begin{center}
\begin{tabular}{c}
\includegraphics[width=8.0cm,angle=0]{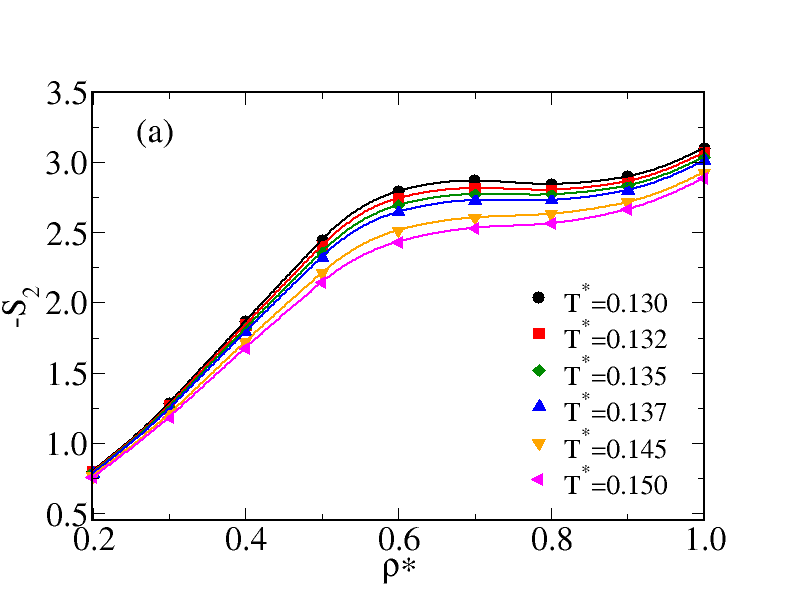} \\
\includegraphics[width=8.0cm,angle=0]{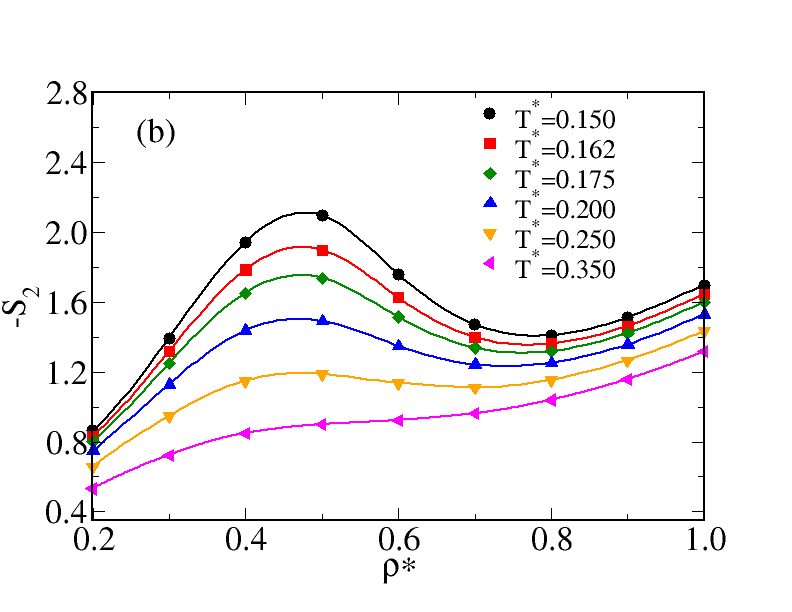} 
\end{tabular}
\caption{Pair translational entropy $S_2$ 
as a function of the density 
along several isotherms for $\lambda=0.1$ and $\alpha=0.6$ (a) and 0.8 (b).
}
\label{fig:s2}
\end{center}
\end{figure}

In order to investigate the fluid structure and to ascertain the presence 
of possible anomalies of this system,
we have carried out standard Monte Carlo (MC) simulations in the NVT ensemble.
We have kept fixed the total number of dimers ($N=864$) enclosed in a cubic
box with periodic boundary conditions. As for the simulation runs, for each
investigated value of $\lambda$ we have first performed a series of simulations
at relatively high temperatures (specifically $T^*=0.40$) upon progressively 
increasing the density from 0.2 to 1.0. 
Once obtained the equilibration, we have gradually 
cooled the system with a $\Delta T$ step of 0.05. For each run we have
performed $2 \times 10^5$ steps to equilibrate the system, then followed by
the same number of steps to collect statistical properties. 
At low temperatures we have extended the number
of steps up to $5 \times 10^5$ in order to ensure a proper equilibration
even in such conditions.

\section{Results}
We first remind that in the monomeric case, corresponding to $\lambda=0$, 
density and structural anomalies have been observed for both $\alpha=0.6$ and
$\alpha=0.8$~\cite{Franz:JPCB,Franz:MolPhys}. 
The density anomaly indicates an unusual expansion of the system upon cooling, 
with the density increasing till to reach a maximum and then decreasing. The
structural anomaly is instead characterized by the unusual behaviour of the 
pair translational entropy $S_2$ defined as~\cite{Green}:
\begin{equation}
S_2/k_{B}=-\frac{1}{2}\rho\int d{\bf r} [g_{cm}(r) {\rm ln}  g_{cm}(r) -g_{cm}(r) +1]
\end{equation}
where $g_{cm}(r)$ is the pair distribution function between the centers of 
mass of two dimers. This quantity effectively characterizes the degree of pair translational
order present in the fluid~\cite{Truskett}. At variance with a simple fluid, $-S_2$ has a
a non-monotonic behaviour for systems 
interacting through a CS potential.
\begin{figure}
\begin{center}
\begin{tabular}{c}
\includegraphics[width=8.0cm,angle=0]{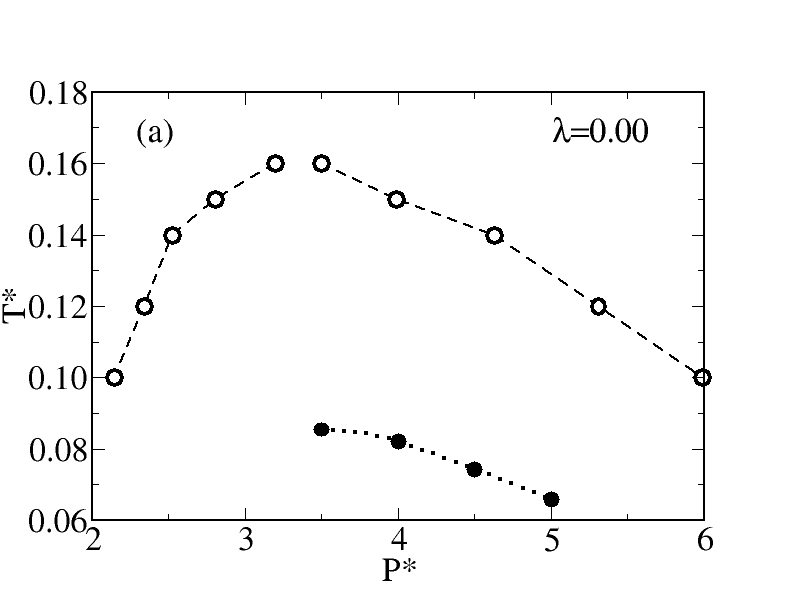} \\
\includegraphics[width=8.0cm,angle=0]{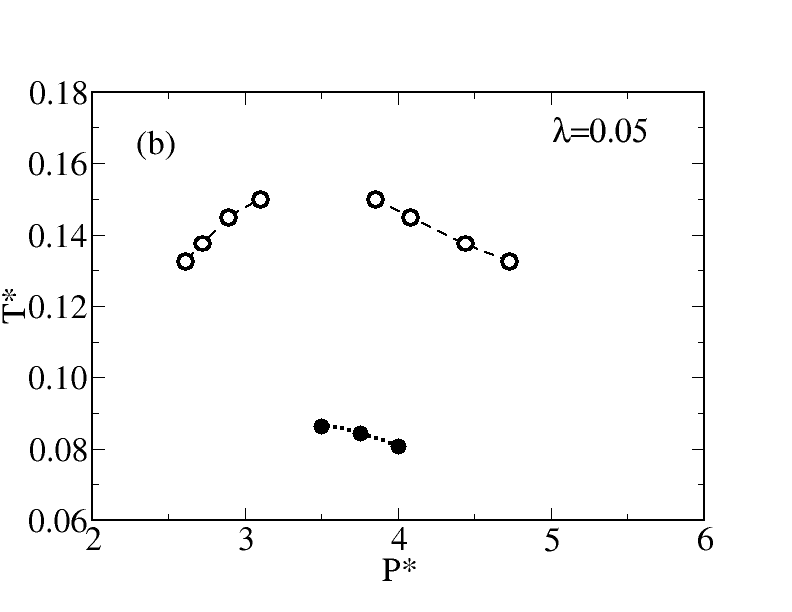} 
\end{tabular}
\caption{Loci of structural (open symbols) and density (full symbols) 
anomalies in the pressure-temperature plane at $\alpha=0.6$ upon increasing 
$\lambda$. } 
\label{fig:a06}
\end{center}
\end{figure}
\begin{figure*}
\begin{center}
\begin{tabular}{cc}
\includegraphics[width=8.0cm,angle=0,height=5.4cm]{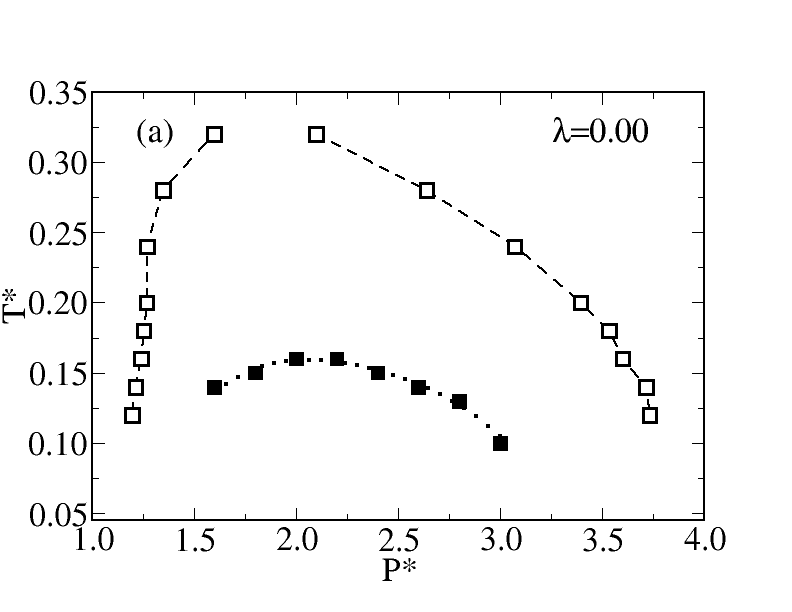} 
\includegraphics[width=8.0cm,angle=0,height=5.4cm]{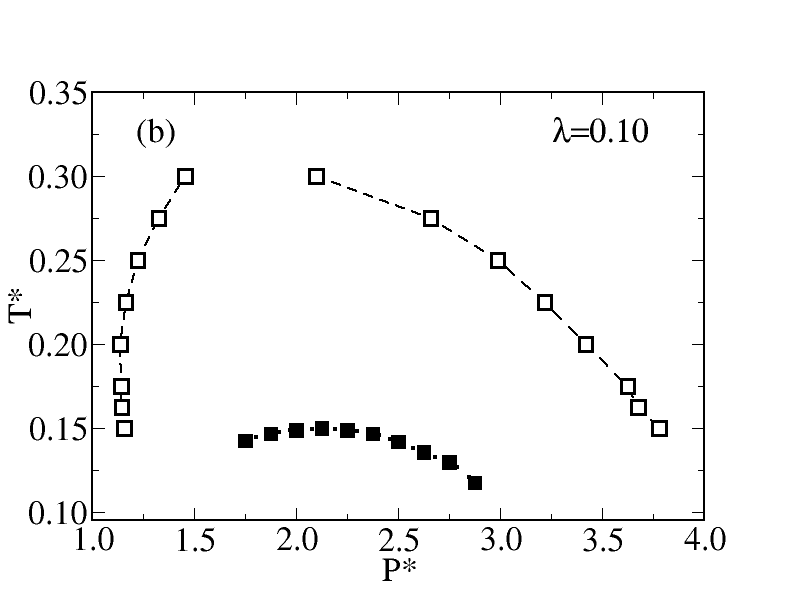} \\  
\includegraphics[width=8.0cm,angle=0,height=5.4cm]{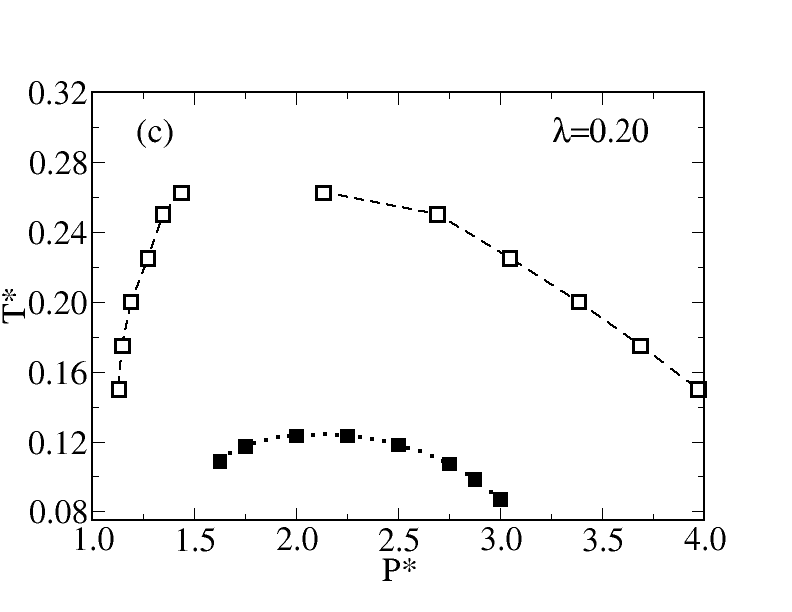} 
\includegraphics[width=8.0cm,angle=0,height=5.4cm]{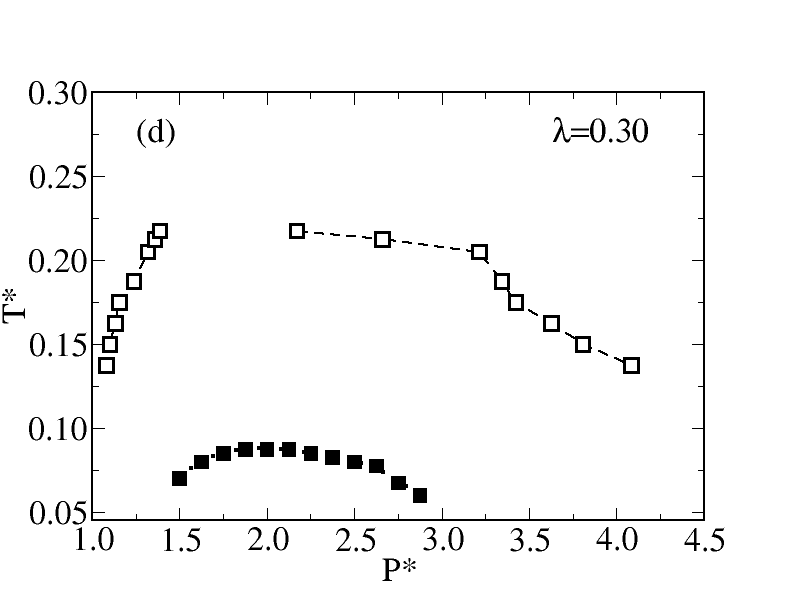} \\
\end{tabular}
\caption{Loci of structural (open symbols) and density (full symbols) 
anomalies in the pressure-temperature plane at $\alpha=0.8$ upon increasing 
$\lambda$. } 
\label{fig:a08}
\end{center}
\end{figure*}

The presence of these anomalies for both the values of $\alpha$ in the 
monomeric case suggests that the
onset of anomalous behaviours in MIP fluids is not stricly connected with the
double length scale of the potential (attained at $\alpha=0.8$) but it may
be found even if the potential exhibits only one length scale (as, for 
instance, at $\alpha=0.6$). However, in the latter case, anomalies are more
difficult to find and very extensive and careful simulations need to be 
performed to locate them. The effect of increasing $\lambda$ is expected to
further influence the structural and 
thermodynamic behaviour of the system; we shall see in 
the follow how the balance between aspect ratio and length scales of the 
potential is responsible of the onset or disappearance of anomalous behaviours
in MIP dimer fluids.
Also, following the prescription suggested by de Oliveira and coworkers
in their previous investigation of waterlike anomalies in dimer 
systems~\cite{Barbosa:10}, pressure and 
temperature shall be rescaled by a factor of 4 if $\lambda \neq 0$. 
This is to guarantee
a better comparison with the monomeric case ($\lambda=0$), where
the effective intermolecular interaction between particles is four times
weaker. 

We report in Fig.~\ref{fig:TMD} the behaviour of the density as a function of
the temperature at a given pressure; more specifically, 
we show results for $\alpha=0.6$, $\lambda=0.05$ and $P^*=3.75$ in panel (a), 
while data for $\alpha=0.8$, $\lambda=0.2$ and $P^*=2.75$ are
reported in panel (b). Both the two curves display a maximum of $\rho^*$ upon
increasing $T^*$, followed by a rapid decay; this scenario testifies 
the presence
of a temperature of maximum density (TMD), 
whose value is strictly dependent on the particular pressure 
considered. We anticipate that the increase of $\lambda$ tends to hinder the
development of this density anomaly; specifically, the higher values of 
$\lambda$ where a TMD is still observed are 0.05 for $\alpha=0.6$ and 0.30 for
$\alpha=0.8$. 

Results for the pair translational entropy 
as a function of the density 
for $\alpha=0.6$ and 0.8 at the same $\lambda=0.1$ are 
reported, respectively, in panels (a) and (b) of Fig.~\ref{fig:s2}.  
At high values of $T^*$, $S_2$ monotonically increases with $\rho^*$ and no
anomalies are observed.
Upon decreasing the temperature, the behaviour of $S_2$ ceases to be monotonic
and shows the presence of a maximum and a minimum more and more structured.
This feature is clear especially for $\alpha=0.8$ (Fig.~\ref{fig:s2}b) but it is
observed, albeit in a less extent, for $\alpha=0.6$ also (Fig.~\ref{fig:s2}a).
The non-monotonic behaviour of $S_2$ is usually related to the presence of 
a structural anomaly in the fluid~\cite{Giaquinta:05}; as a consequence, such a kind of anomaly, 
already observed for both $\alpha=0.6$ and $\alpha=0.8$ in the monomeric case,
survives for $\alpha=0.6$ even if $\lambda=0.1$, where the density anomaly is
disappeared. 

The specific values of maxima and minima define a portion of the
temperature-pressure plane corresponding to the structural anomaly region.
At the same time, values attained by TMD define, in the same plane, the 
density anomaly region. In Fig.~\ref{fig:a06} we show these two regions for 
$\alpha=0.6$ and $\lambda=0$ (a) and $\lambda=0.05$ (b). 
As visible, the small increase of $\lambda$ is enough to shrink
both structural and density anomaly regions. Actually, a maximum and a minimum 
in $S_2$ can be hardly detected for $\lambda=0.05$ if $T^* < 0.13$; this is
due to the development of solid-like features in $g_{cm}(r)$ that cause a sharp 
increase of the correspondig pair translational entropy. 
As a consequence, the profile of
$S_2$ vs $\rho^*$ shows discontinuities preventing the detection of maxima
and minima in these functions. These circumstances suggest that the shrink
of anomalies regions may be ascribed to an increase of the solid region with
respect to the monomeric case.   

Regions of structural and density anomalies for $\alpha=0.8$ at various
$\lambda$ are reported in Fig.~\ref{fig:a08}. The effect of increasing the
aspect ratio is now reflected in the lowering of the temperature where the
anomalies are observed. That is, the dimer system needs to be
more and more cooled before showing anomalous behaviour; if $\lambda > 0.3$,
the density anomaly is no more observed, whereas the structural anomaly, 
which develops at higher temperature, is still found.
\begin{figure}
\begin{center}
\begin{tabular}{c}
\includegraphics[width=8.0cm,angle=0]{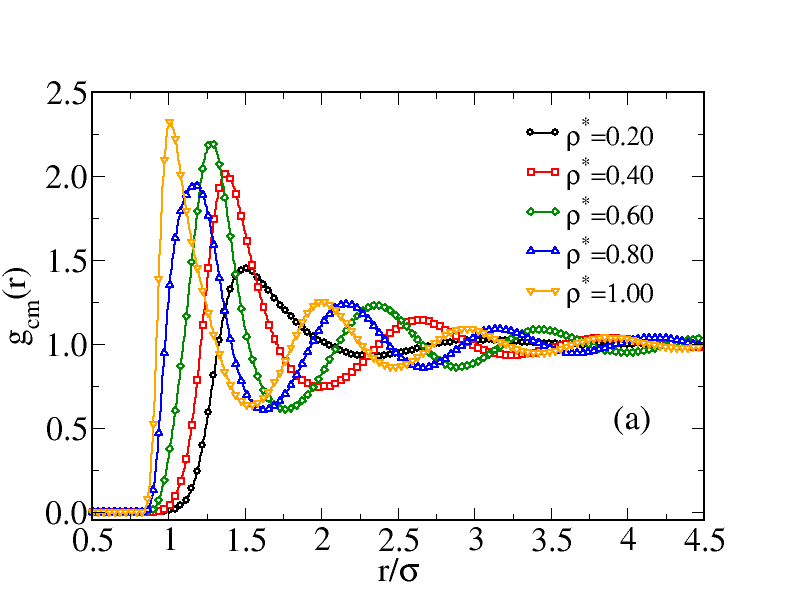} \\
\includegraphics[width=8.0cm,angle=0]{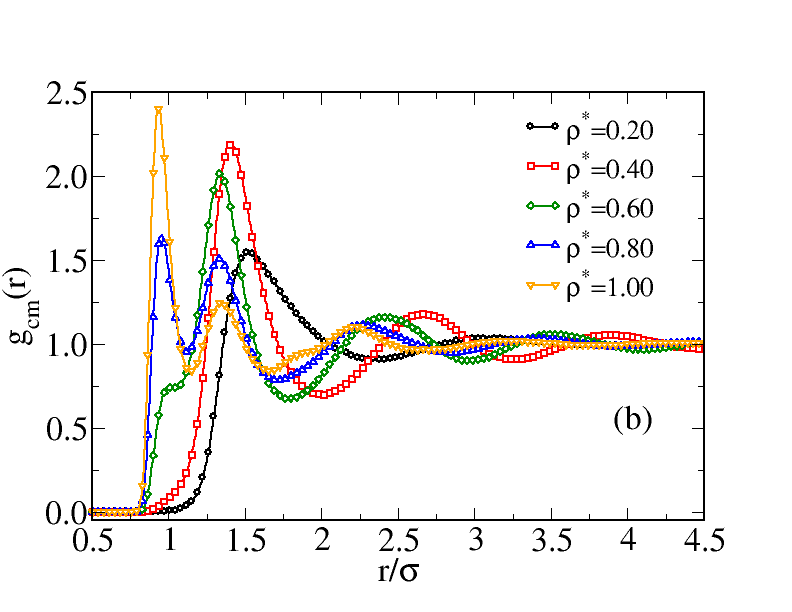} 
\end{tabular}
\caption{Radial distribution functions between the centers of 
mass at $T^*=0.15$ for $\alpha=0.6$ and 
$\lambda=0.05$ (a) and for $\alpha=0.8$ and $\lambda=0.10$ (b)
upon increasing $\rho^*$.} 
\label{fig:gr}
\end{center}
\end{figure}

Finally, the fluid structure of MIP dimers is investigated in 
Fig.~\ref{fig:gr}, where results for $g_{cm}(r)$ as a function of the density 
for $T^*=0.15$ and $\alpha=0.6$ (a) and 0.8 (b)
are reported. 
When the potential exhibits only one length scale (panel (a)), 
the progressive increase of $\rho^*$ causes the shift of the
first peak of $g_{cm}(r)$ towards low values of $r/\sigma$. This is due to the lower
typical distance between two dimers ($\lambda=0.05$ in this case) 
when the system is compressed. A different behaviour is observed for 
$\alpha=0.8$ and $\lambda=0.1$ (panel (b)): 
now, a double peak is clearly observed in
$g_{cm}(r)$ if $\rho^* \geq 0.60$. This feature is due to the presence of a
double length scale in the potential, actually attained for $\alpha=0.8$. 
In this case, two distinct arrangements between dimers become possible, 
with the one corresponding to the first peak in $g_{cm}(r)$ more favored upon
increasing the density. However, despite the differences in the
$g_{cm}(r)$, structural and density anomalies are both observed in the two 
thermodynamic state points considered in Fig.~\ref{fig:gr} (see 
Fig~\ref{fig:a06} and Fig~\ref{fig:a08}).

The overall behaviour of these anomalies as a function of
$\alpha$ and $\lambda$ is schematically summarized in Table~\ref{tab:anom}:
in particular, the effect of progressively increase the aspect ratio when the 
potential has one or two length scale is enlightened, at the same time
enhancing the persistence of the structural anomaly when TMD disappears.

\begin{table}[t!]
\caption{Overall behaviour of density and structural anomalies upon 
varying $\alpha$ and $\lambda$.}
\label{tab:anom}
\begin{center}
\begin{tabular*}{0.48\textwidth}{@{\extracolsep{\fill}}cccc}
\hline\hline
$\alpha$ & $\lambda$ & TMD & Structural Anomaly \\
\hline
 0.6 & 0.00 & \cmark & \cmark \\
 0.6 & 0.05 & \cmark & \cmark \\
 0.6 & 0.10 & \xmark & \cmark \\
 0.8 & 0.00 & \cmark & \cmark \\
 0.8 & 0.10 & \cmark & \cmark \\
 0.8 & 0.20 & \cmark & \cmark \\
 0.8 & 0.30 & \cmark & \cmark \\
 0.8 & 0.40 & \xmark & \cmark \\
\hline\hline
\end{tabular*}
\end{center}
\end{table}
\section{Conclusions}
We have investigated the fluid phase of a model of dimers interacting through
a core-softened potential 
by means of Monte Carlo simulations with a 
particular focus on their density and structural anomalies. 
Specifically, two dimers interact via a modified inverse-power potential
(MIP), {\rm i.e.} a repulsive potential of inverse-power form modified to
soften the repulsion strength in a range of distances.
We have considered two different conditions, corresponding to one 
($\alpha=0.6$) and two ($\alpha=0.8$) length scales of the intermolecular
potential, analyzing how the increase of the aspect ratio $\lambda$ of the
dimers influences their anomalous behaviour. In the spherically-symmetric
case (i.e. $\lambda=0$) density and structural anomalies have been observed
in both cases already in previous works~\cite{Franz:JPCB,Franz:MolPhys}: here
we have found that a small increase of $\lambda$ (from 0 to 0.05) allows for 
the persistence of such anomalies at $\alpha=0.6$. Upon further increasing 
$\lambda$, the density anomaly disappears, whereas the structural anomaly
is still found for $\lambda=0.10$. As for the two-length scale case, both 
anomalies survive till to $\lambda=0.30$; beyond this value, no density 
anomaly is observed, with the structural anomaly persisting for $\lambda=0.40$.

The fluid structure of MIP dimers has been investigated through the radial
distribution function $g_{cm}(r)$ that shows different behaviours for the two
investigated values of $\alpha$. In particular, the development of a 
double-peak structure for $\alpha=0.8$ is strongly reminiscent of the
two-length scale of the intermolecular potential. Such a feature is not
observed for $\alpha=0.6$, even if density and structural anomalies are
present. These findings suggest that the development of such anomalies in
CS systems is not strictly dependent on the double length scale of the 
potential; moreover the increase of the aspect ratio of the dimers tends 
to suppress the anomalies in a way depending on the specific combination of 
$\alpha$ and $\lambda$. 

Our results pave the way for a more general investigation of CS non-spherical
systems, enlightening the role played by the anisotropy in affecting 
their phase behaviour.  
Such systems may include, for instance, elongated molecules, 
polymers and colloidal dimers, whose experimental 
realization~\cite{Bon:14,Bon:SM} along with theoretical 
models~\cite{Munao:14,Bordin:15} are currently object of increasing interest.


\providecommand*{\mcitethebibliography}{\thebibliography}
\csname @ifundefined\endcsname{endmcitethebibliography}
{\let\endmcitethebibliography\endthebibliography}{}

\end{document}